# Studies on tuning surface electronic properties of hydrogenated diamond by oxygen functionalization

N. Mohasin Sulthana [a,b], K. Ganesan [a,b,1], P.K. Ajikumar [b], and S. Dhara [a,b]

[a] *Homi Bhabha National Institute, Training School Complex, Anushaktinagar, Mumbai – 400094, India.*

[b] *Surface and Nanoscience Division, Materials Science Group, Indira Gandhi Centre for Atomic Research, Kalpakkam - 603102, India.*

## Abstract

Ultra-wide bandgap and the absence of shallow dopants are the major challenges in realizing diamond based electronics. However, the surface functionalization offers an excellent alternative to tune electronic structure of diamonds. Herein, we report on tuning the surface electronic properties of hydrogenated polycrystalline diamond films through oxygen functionalization. The hydrogenated diamond (HD) surface transforms from hydrophobic to hydrophilic nature and the sheet resistance increases from ~ 8 kΩ/sq. to over 10 GΩ/sq. with progressive ozonation. The conductive atomic force microscopic (c-AFM) studies reveal preferential higher current conduction on selective grain interiors (GIs) than that of grain boundaries confirming the surface charge transfer doping on these HDs. In addition, the local current conduction is also found to be much higher on (111) planes as compared to (100) planes on pristine and marginally O-terminated HD. However, there is no current flow on the fully O-terminated diamond (OD) surface. Further, X-ray photoelectron spectroscopic (XPS) studies reveal a redshift in binding energy (BE) of C1*s* on pristine and marginally O-terminated HD surfaces indicating surface band bending whilst the BE shifts to higher energy for OD. Moreover, XPS analysis also corroborate c-AFM study for the possible charge transfer doping mechanism on the diamond films which results in high current conduction on GIs of pristine and partially O-terminated HDs.

Keywords : Hydrogenated diamond; Surface conductivity; Raman spectroscopy; X-ray photoelectron spectroscopy; Conductive atomic force microscopy

---

[1] Corresponding author. Email : kganesan@igcar.gov.in ( K. Ganesan)

# 1. Introduction

Diamond exhibits the most attractive properties such as high thermal conductivity (22 Wcm$^{-1}$K$^{-1}$), ultra-wide bandgap (5.47 eV), high carrier mobility (3800 cm$^2$V$^{-1}$s$^{-1}$ for holes), high electric breakdown field (10 MVcm$^{-1}$), low dielectric constant (~ 5.7), chemical inertness, biocompatibility and wide potential window for electrochemical analysis, which make it as a promising material for electronic devices [1,2]. However, the high intrinsic resistivity and the absence of shallow *n/p*-type dopants impose several difficulties in conventional electronic device fabrication. Hence, the modification of the band structure by alternative methods has a fundamental importance for future applications, such as light emitting devices, thermionic electron emission and high power & high temperature electronics [3]. One simple and effective way of modifying the band structure of diamond is to terminate the surface with functional groups which strongly controls the surface electronic properties such as work function, electron affinity (EA) and surface conductivity (SC). It is well known that the surface electronic properties of diamond are significantly influenced by different types of functionalization such as hydrogen, oxygen, fluorine, nitrogen and amine group [3–5].

The hydrogen (H) termination on diamond is the most successful method to achieve high SC in diamonds. Though the high SC was observed a long ago on the diamonds that were polished under organic solvents, the origin of SC is understood only recently through experimental and theoretical modelling [6–10]. One of the proposed models to explain the high SC of diamonds is the surface transfer doping mechanism which allows to form a two dimensional hole gas (2DHG) on surface and sub-surface of hydrogenated diamond (HD) [6–8,11]. The negative electron affinity (NEA) of HD is understood to be a prerequisite for such a high SC with upward band bending [9]. In contrast, oxygen (O-) terminated diamond (OD) surface possesses positive electron affinity (PEA) which leads to insulating surface with downward band bending [12,13]. Thus, the HDs and ODs have contrasting electronic properties. Moreover, the electronic properties of the functionalized diamond can be tuned from highly conductive surface to insulating surface by controlling the H- and O- concentration on the diamond surface. Importantly, the surface electronic properties also play a significant role in controlling the chemical reactivity and catalytic performance on the diamond surface.

The O-termination on diamonds is commonly achieved by various techniques such as oxygen plasma treatment, chemical oxidation under acid medium, anodic polarization in an alkaline electrolyte and UV ozone treatments [9,12–16]. Among them, UV ozone treatment is a simple and an efficient method for O-termination of diamond and also, it has advantages of controlled oxidation only on the ozone exposed area. The OD surface can have variety of bonding states such as singly oxidized (C-O-H, C-O-C, C-O-O-C), doubly (C=O, O-C-O) oxidized or combination of them [14]. Though the ether C-O-C and ketone C=O bonds are mostly reported in experimental, the calculations indicate the hydroxyl C-OH and ketone C=O states are stable for high oxygen coverage on surface [12,15–17]. Further, the surface electronic properties of functionalized diamonds are very sensitive to the nature of chemical bonding states. Hence, the knowledge on surface configuration is essential for different coverage of H- & O- atoms on diamonds. Although intensive X-ray photoelectron spectroscopy (XPS) studies were carried out on HDs and ODs for understanding the surface electronic structure in terms of band bending, 2DHG formation, and work function [3,10] the studies on partially oxygenated diamond surfaces are limited. Such partial ozonation would allow to tune of surface electronic properties of diamonds that will help to control the electronic device characteristics of HDs for desired applications in the areas of field effect transistors, electrochemical devices, and chemical gas and bio-sensors [13,18]. For example, the HD based MOSFETs often have normally-ON state due to its high surface conductivity. But, it has been demonstrated that a partial C-O channel makes the device to be normally-OFF state with high breakdown voltage (~2000 V) [13]. Also, the surface carrier density of diamonds is shown to be tuned by controlling the H-termination process conditions and the variation in carrier density can also provide the opportunity to tune to the drain voltage and drain currents in HD MOSFETs [18]. However, the studies on the structural and physical properties that evolve with partial oxygenation on diamond surface are scarce. This motivates us to undertake a detailed study on the variation of electronic properties of partially ozonated HDs.

In this study, we endeavor to tune the SC of HD surface in well controlled manner using ozonation process. This study demonstrates a significant role of O- termination on diamond to manipulate the surface electronic structure. The H- and O- terminations on diamond surface was studied qualitatively using wetting contact angle (WCA) and electrical measurements. In addition, the local current measurements were also carried out using conductive atomic force microscopy (c-AFM). Further, a preliminary study on the variation in surface electronic structure of the pristine and ozonated HD films are monitored using XPS. Furthermore, the effect of surface charge density on the Raman spectroscopy and fluorescence emission of SiV$^-$ color centers were investigated.

## 2. Experimental methods

Microcrystalline diamond films were grown on SiO$_2$/Si by hot filament chemical vapor deposition using methane and hydrogen as feedstocks in the ratio of 2 : 200, as discussed earlier [19,20]. Subsequent to growth, the in-situ hydrogen termination on diamond surface was achieved by admitting ultrahigh pure hydrogen into the chamber at a flow rate of 200 sccm for 1200 s under working pressure of ~ 40 mbar. During hydrogen treatment, the substrate and filament temperatures were maintained at 800 and ~ 2000 ºC, respectively. After H termination, the samples were allowed to cool down to room temperature in hydrogen atmosphere. The H-terminated samples were exposed to air atmosphere for 2 days to reach surface charge transfer equilibrium state. Later, the HD films were exposed to ozone atmosphere for different durations of 30, 60, 90, 300 and 1200 s using UV/ozone pro cleaner (UV ozone cleaner, Ossila). Here, the ozone is produced by illuminating atmospheric air with high intensity ultra-violet light source in the reactor chamber. Since the UV/ozone cleaner works at ambient condition with atmospheric air, there is a possibility of contamination of diamond surface with atmospheric impurities and residual NO$_2$ molecules than can be generated along with ozone, during ozonation process. In addition, the partial O- terminated samples were also exposed to air atmosphere for at least two days to reach an equilibrium state before considering for experimental analysis.

A commercial AFM (M/s NT-MDT, Russia) was used to examine the variation in microstructure of the samples. The effect of H- / O- functionalization on the wetting characteristics of diamond surface was studied by water contact angle measurement using telescope-goniometer (APEX, ACAM D1N01, India). Liquid drop analysis (LB-ADSA) was performed using ImageJ software to estimate the WCA. Macroscopic electrical measurements were carried out with Agilent B2902A precision Source/measure unit using four probe measurements with tungsten whisker probes with tip radius of curvature of ~ 20 μm and the dimension of the samples are ~ 2 mm x 2 mm. Raman and photoluminescence (PL) spectra were recorded using micro-Raman spectrometer (In-via, Renishaw, UK) having diode laser with wavelength of 532 nm and a grating of 1800 grooves/mm. In addition, the local electrical properties of the HD and ozonated surfaces were also measured by c-AFM at nanometer-scale resolution. Pt cantilevers with stiffness constant of ~ 1 Nm$^{-1}$ was used for the study. The c-AFM studies were performed under controlled environment with relative humidity of < 2 % by gently blowing dry nitrogen to avoid the tip oxidation during current measurements. Further, a low current preamplifier was also used to measure the local current down to 200 fA for 300 and 1200 s ozonated HD samples since the resistance of these sample are very high ( > 10 GΩ). In addition, a qualitative comparison of surface chemical properties of the samples were also studied using XPS (M/s SPECS, Germany) system equipped with a monochromatic Al Kα source (1486.6 eV) operated at 350 W with detection pass energy of 10 eV.

## 3. Results

### 3.1. Surface topography

Figure 1 shows our AFM topography and simultaneously measured friction force mapping of HD surface. The diamond film exhibits multiple faceted surface morphology with thickness of ~ 2 μm. Here, the diamond film is grown under constrained environment imposed by the coalescence of multiple diamond nucleation and growth in lateral and vertical direction. In general, the texture and morphology of the diamond films are controlled by the displacement rate of the (100) and (111) planes [21]. As can be seen in Fig. 1a, the morphology of the grown diamond film can be expressed as <110> texture associated with (100) and (111) planes since the diamond films are grown under low the $CH_4$ concentration at relatively high substrate temperature [21]. Here, the (100) and (111) planes are indicated as triangle (red) and square (green) symbols respectively, on the friction image given in Fig. 1b. Further, the (111) facet displays higher surface roughness than that of (100) surface and it can be clearly visualized in the friction image which shows distinct contrast for smooth and rough surfaces, as shown in Fig. 1b. Thus, the roughness can be used as a tool to differentiate the (100) and (111) facets.

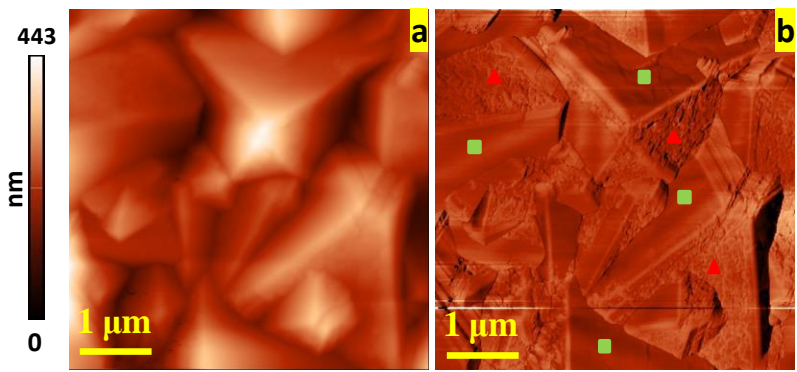

Fig 1. Simultaneously measured (a) AFM topography and (b) frictional force mapping of H-diamond surface. The triangle (red) and square (green) symbols in the images indicate the (111) and (100) planes respectively, on diamond surface.

**3.2. Wetting contact angle analysis**

Figure 2 depicts the variation of water WCA which decreases from 103 down to 33º as a function of ozonation duration of HD surface. The corresponding optical images of the WCA are also given as inset in Fig. 2. The measured WCA of 103º degrees of as-prepared HD confirms the hydrophobic nature of the surface which is in good agreement with the reported data [22]. This hydrophobic nature is attributed to low surface free energy of HD surface. Also, only a very weak van der Waals interaction exists between fully H-terminated diamond surface and water molecules that makes the surface to be hydrophobic [4]. However, when the surface consists of oxygen functional groups, the surface becomes polar and interacts with other polar medium like water [5]. Thus, the surface becomes hydrophilic with high surface free energy. As shown in Fig. 2, after ozonation for 30 s, WCA immediately decreases down to 74º. Further, WCA continues to decrease gradually down to 32º for 1200 s ozonation and the surface becomes hydrophilic.

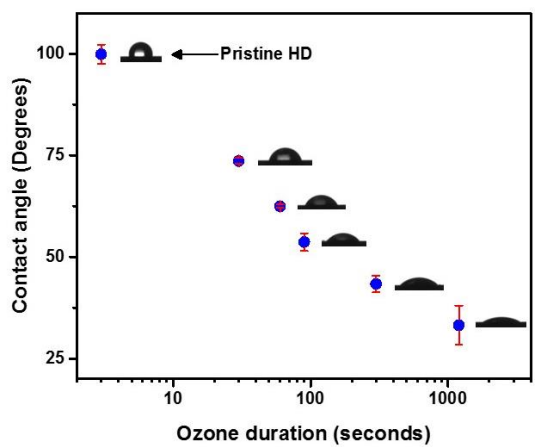

Fig. 2. The variation of wetting contact angle of hydrogenated diamond as a function of ozonation duration. The inset images represent the corresponding optical images of the wetting water contact angle.

Here, we note that WCA didn't vary significantly for the samples exposed to ozone for 1200, 1500 and 1800 s. Hence, we presume that the diamond surface attains O- saturation limit after the ozone exposure for 1200 s. Moreover, the gradual transformation of wettability can be correlated to the systematic substitution of surface hydrogen by oxygen functional groups during ozonation.

### 3.3. Macroscopic electrical properties

Figure 3a displays the variation of sheet resistance of HD surface as a function of ozonation duration from 0 to 1200 s. The sheet resistance of the as-prepared HD surface is found to be 7.6 x $10^3$ Ω/sq. Further, the sheet resistance increases exponentially with ozonation and it reaches to ~ 6.5 x $10^7$ Ω/sq. for 300 s ozonation. The sheet resistance of the fully ozonated diamond for 1200 s has increased to beyond > 1 x $10^{10}$ Ω/sq., which is the measuring limit of the equipment. Fig. 3b depicts the variation of hole density and mobility of pristine and ozonated HDs for 30, 60 and 90 s. Note that the ozonation of HD films for a short duration of 90 s decreases the surface hole density by two orders from 1.05 x $10^{13}$ to 2.49 x $10^{11}$ cm$^{-2}$. However, the mobility does not vary significantly for the partially ozonated HDs. Further, the measured hole density and mobility of the diamond films are in good agreement with the reported values for polycrystalline diamond films [23]. Here, the low sheet resistance and high hole density of pristine and partially ozonated HDs confirm the presence of hole accumulation layer which occurs through surface transfer doping mechanism [6–9]. Further, the increase of O- concentration on diamond surface, as inferred from WCA analysis, the sheet resistance ( hole density ) increases (decreases) with ozonation duration.

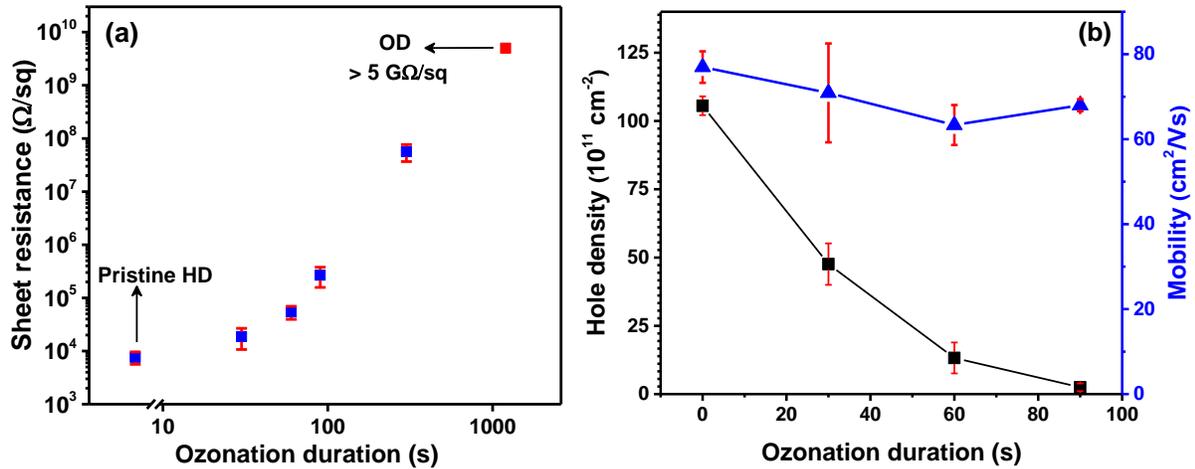

Fig. 3. The variation of (a) sheet resistance, and (b) carrier density and mobility of HD surface with ozonation duration.

### 3.4. Conductive AFM studies

Figure 4 shows the simultaneously measured topography, friction and current images of the pristine and ozonated HD samples for 60, 90, 300 and 1200 s, measured at an applied bias of 0.15, 0.33, 0.90, 1.00, 1.30 and 20.00 V, respectively. As shown in Fig. 4(a-e), the topography of the films display majority of mixed (100) and (111) planes along with several additional orientations near the edges of these planes and also, secondary nucleation of diamond grains on the surface. The corresponding friction images [Figs. 4(f-j)] clearly reveal the distinct nature of the different planes, edges, steps and secondary nucleation with sharp contrast variation. Here, the triangle (red) and square (green) symbols are used to indicate the (100) and (111) planes respectively, as shown in friction images given in Figs. 4(f-j). As can be evidenced from the current mappings in Figs. 4(k-n), the majority of grain interiors (GIs) have higher current than that of the grain boundaries (GBs), even though the latter one has a large amount of $sp^2$-rich C-C bonding which are known to have high conductivity [24]. Further, a certain edges / planes between the major crystallographic planes also display high current which are shown in dashed rectangle in Figs. 4(k,l). The observation of high SC on GIs reconfirms the presence of high density of charge carriers on diamond surface due to transfer doping mechanism induced by hydrogenation [11,25,26]. Further, we note here that the current conduction on GIs is highly inhomogeneous. Especially the current flow on (111) planes is in the order of a few nanoamperes (nA) which is much higher as compared to (100) planes wherein the current flow is ~ 100 picoamperes (pA) on the pristine and 60 s ozonated HD samples. On the other hand, the current flow appears to be marginally higher on (100) planes as compared to (111) planes as can be evidenced in Figs. 4(m,n), for the HDs ozonated for 90 and 300 s. However, there is no measurable current flow on GIs and GBs of OD films that are ozonated for higher than 300 s even at higher applied bias of 20 V and a typical current map of OD (ozonated for 1200 s) is shown in Fig. 4o. This lowest conductance confirms the complete removal of surface charge transfer doping of diamond after ozonation for 1200 s. Moreover, the average maximum current measured on (111) facet decreases drastically from ~ 5000 to ~ 5 pA while it decreases gradually from ~ 100 to 10 pA on (100) facet when the pristine HD film is ozonated upto 300 s (Fig. 4). The drastic reduction in surface current reveals that the oxidation rate is higher on (111) facets than that of the (100) facets. This result is consistent with the oxidation rate of polycrystalline HD (111) and (100) facets as evaluated based on electron energy loss spectroscopy [27].

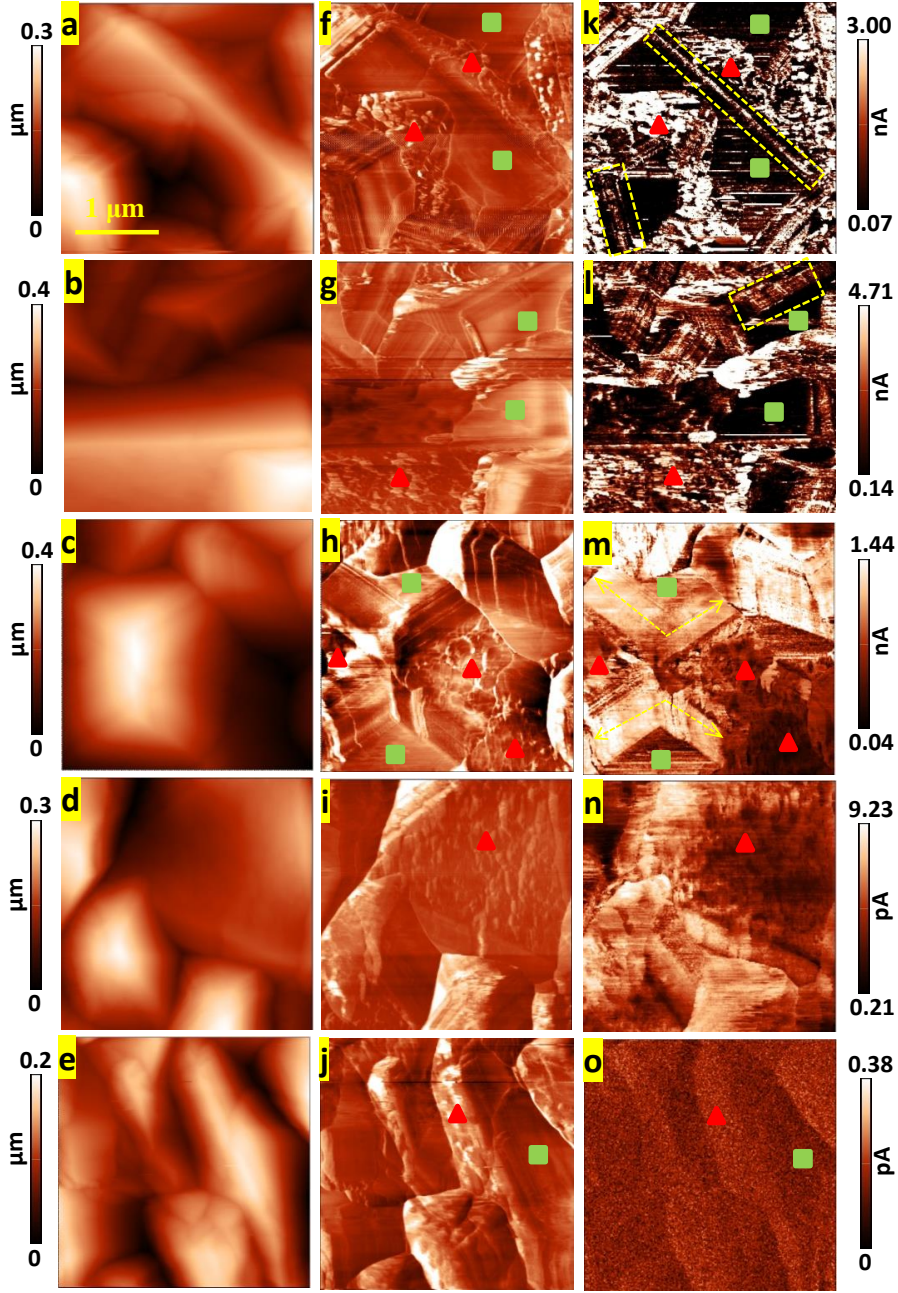

Fig. 4. Simultaneously measured (a-e) topography, (f-j) friction force and (k-o) current images of hydrogenated diamond (a,f, k, ), and ozonated diamond surfaces for 60 s (b, g, l), 90 s (c, h, m), 300 s (d, i, n) and 1200 s (e, j o). The scale bar is same for all the images.

Based on the c-AFM current measurements, the local resistance is estimated to be ~ 7.02 x $10^7$, 1.49 x $10^8$, 6.94x$10^8$ and 1.63 x$10^{11}$ Ω for the samples HD, and 60, 90 and 300 s ozonated HD, respectively. These values are calculated based on the maximum current value from the c-AFM mapping at an applied bias of 0.15, 0.33, 0.90, and 1.30 V, respectively. Note, a 10 MΩ resistance was also connected in series in order to limit the maximum current flow through the AFM tip. Here, the local resistance value measured by c-AFM also increases with ozonation duration. In addition, the net current conduction on these diamond surface can also be expressed as random resistor network model. Here, the local surface conductance of diamond facets like (111) & (100) planes, grain boundaries and the tunneling barrier between the diamond facets play significant role in determining the net surface conductivity. Further, the above mentioned resistor elements get affected by partial ozonation eg. the conductivity of diamond facets decreases while it increases on grain boundaries with ozonation. Depending upon the random resistor networks, there is a probability to form maximum conductive path on surface at macroscopic scale and this leads to the difference in surface conductivity measured by four probe measurement and c-AFM technique. Hence, the macroscopic electrical data is used as a representative electrical properties of the diamond films while the c-AFM data is used only for comparison of resistance among adjacent diamond grains.

**3.5 Raman and fluorescence spectroscopy**

The PL spectra recorded on a selective grain of 0, 30, 60 and 1200 s ozonated HDs (labeled as S1, S2, S3 and S4, respectively) are shown in Fig. 5a. The PL spectra were repeated sequentially on a few selected grains of the same sample through position marking and in each step, the sample was ozonated for specific duration before the spectroscopic measurements. The position marked spectra help to avoid the ambiguity in sample inhomogeneity due to polycrystalline nature. Further, these spectra were recorded under identical experimental parameters for direct comparison of intensity. In Fig. 5a, the sharp peak at ~ 573 nm (1332 $cm^{-1}$) and the broad peak at ~ 738.5 nm are assigned to the Raman band and the fluorescence (FL) emission from $SiV^-$ color centers in diamond lattice, respectively. Here, the incorporation of $SiV^-$ centers in diamond lattice is result of Si substrate which acts as source during HFCVD growth. Typical full width at half maximum of diamond Raman band is ~ 5 $cm^{-1}$ and does not vary significantly with ozonation process. The narrow FWHM and absence of Raman signature associated with non-diamond bands clearly indicate the high structural quality of the diamond films [19]. The magnified part of the Raman and FL spectra from a single grain of the pristine and ozonated HDs are shown in Figs. 5b and 5c, respectively. As shown in Fig. 5d (top panel), the absolute peak intensity of the diamond Raman band (1332 cm$^{-1}$) increases with ozonation duration for 30 and 60 s. However, the sample S4 shows lower intensity as compared to that of sample S3. The increase in Raman intensity is attributed to the decrease in screening effect due to the reduction of surface charge carrier density by removal of hydrogen. On the other hand, the SiV$^-$ fluorescence emission intensity was inhomogenous with respect to position and hence, the intensity of SiV$^-$ emission was first normalized with Raman intensity since they share the common lattice structure. Later, the change in the intensity (with respect to pristine HD sample, S1) of SiV$^-$ emission is calculated as a function of ozonation duration, as shown in plot Fig. 5d (lower pannel) ie. $\{I_{s1} + [(I_s-I_{s1})/I_{s1}]\}(\%)$. Here, the intensity of $I_{s1}$ (pristine HD) is considered as 100% for easy comparison. This increase in FL emission intensity is associated with increase in number density of SiV$^-$ color centers in ozonated diamond films [28].

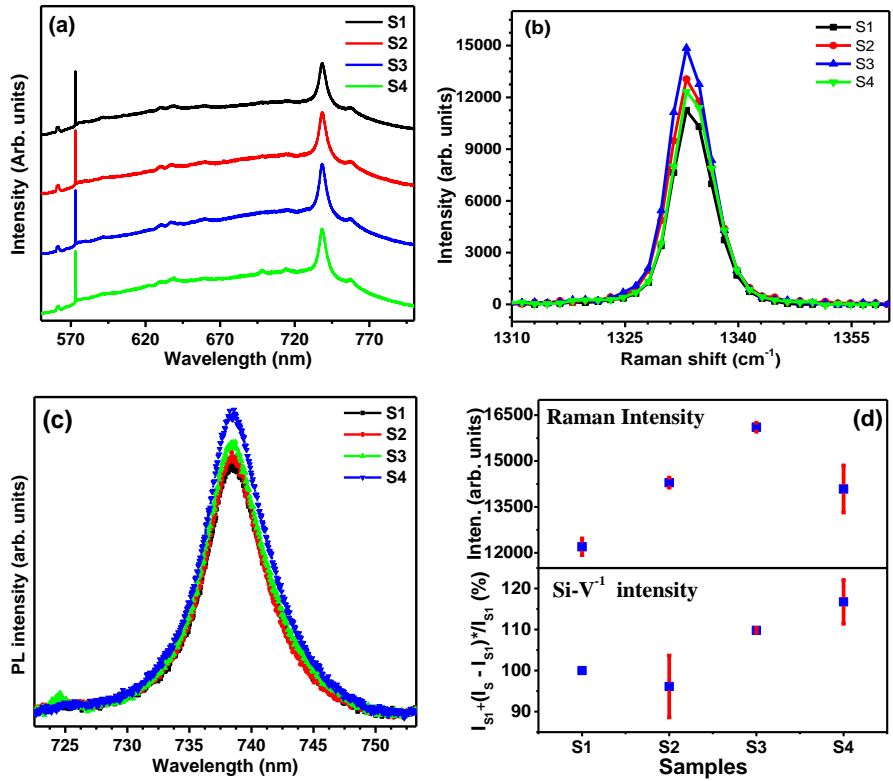

Fig. 5. (a) Photoluminescence spectra of HD and ozonated HD films recorded on a selected grain under the excitation wavelength of 532 nm. These spectra are stacked vertically without background subtraction for clarity. The corresponding absolute intensity comparison of (b) Raman and (c) SiV$^-$ fluorescence emission spectra with suitable background subtraction. (d) The statistical variation of peak intensity of the Raman band (upper panel) and normalized fluorescence emission intensity of SiV$^-$ color centers (lower panel) with respect to sample S1, under identical experimental conditions.

**3.6. X-ray photoelectron spectroscopy**

Figure 6 shows the C1$s$ spectra of pristine and ozonanted HD films (S1-S4) with deconvolution for different bonding configurations. Here, the C1$s$ peak is fitted to Voigt function with 85 % Gaussian and 15 % Lorentzian character, after subtracting Shirley background and the best fit parameters are given in Table 1. It is known that the angle dependent XPS is the most valuable for evaluating true surface electronic properties including O quantification in terms of ML. Since the conventional XPS measurements have several limitations, we limit our XPS analysis only to a qualitative comparison between the surfaces of pristine and ozonated H-diamonds. Moreover, the polycrystalline diamond films have several facets with different angle of orientation to the surface normal. Thus, the bulk sensitive mode XPS (the electron polar angle of q=0 to the surface normal) experiment is not completely the 0º angle orientated experiment. Hence, still we believe that the bulk sensitive mode XPS can also be used to some extent for understanding these samples and to get the surface information qualitatively.

The XPS spectra exhibit a dominant peak (P1) at the binding energy (BE) of 284.4, 284.5, 284.6 and 285.0 eV which are assigned to the C-C $sp^3$ bonds in the bulk diamond for the samples S1, S2, S3 and S4 respectively, as shown in Fig. 6. This peak undergoes a redshift in BE for the samples S1, S2 and S3, as compared to hydrogen free diamond C-C $sp^3$ bond of ~ 285.1 eV. Since hydrogen termination on diamond causes the upward band bending due to surface Fermi level pinning, the BE of the C-C $sp^3$ bonding in HDs is reduced [29,30]. In partially / fully ozonated HD samples (S2-S4), the $E_F$- $E_V$ increases due to the local PEA which decreases the surface charge transfer doping process. Consequently, the surface hole density decreases significantly leading to the reduction in upward band bending. Hence, C-C $sp^3$ peak shifts slowly to higher BE for 30 and 60 s ozonated samples. When HD surface is fully oxidized, the surface exhibits a downward band bending due to the difference in surface and bulk electronic states caused by reconstruction of surface with oxygen. This downward band bending further shifts the C-C $sp^3$ peak to higher BE in fully ozonated sample. The reported values of C-C $sp^3$ peak shifts for hydrogenated and ozonated surfaces from the flat band surface are 0.4 and 0.2 eV, respectively [14]. Hence, the observed peak shift of 0.6 eV between HD and 1200 s ozonated surface is consistent with the reported values [14].

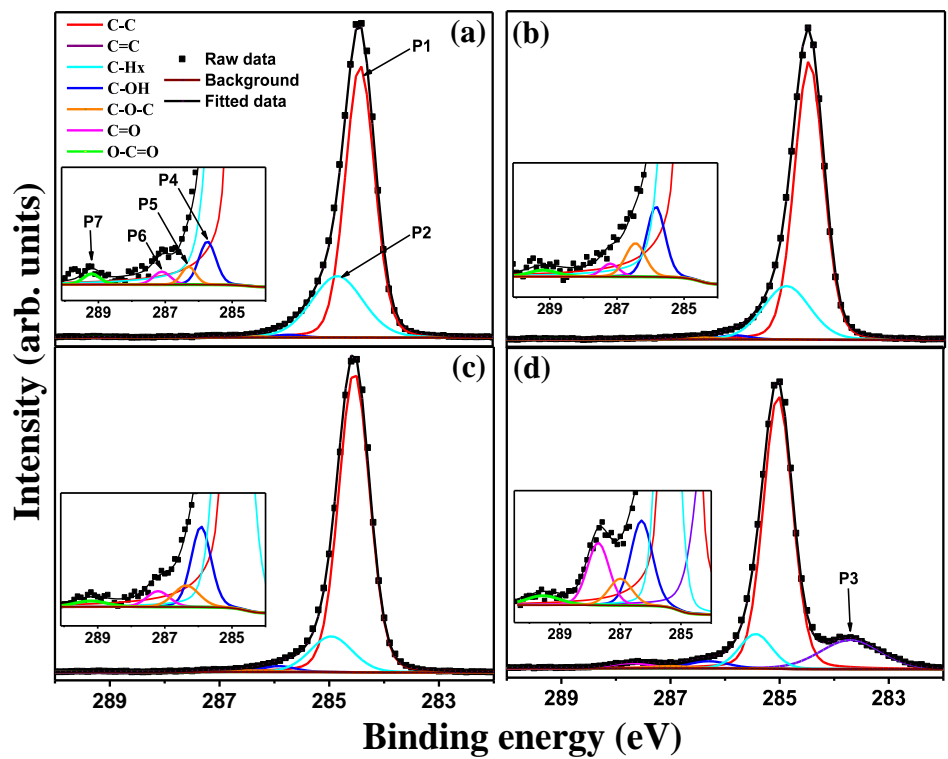

Fig 6. XPS C1*s* core level spectra and its deconvoluated profiles for (a) pristine H-diamond and ozonated for (b) 30 s, (c) 60 s and (d) 1200 s. The inset in each figure represents a magnified portion of the respective C1*s* spectrum.

Table 1. The parameters extracted from XPS analysis for HD and ozonated diamond films

| Sample | Peak 1 C-C | | Peak 2 C-Hx | | Peak 3 C=C | | Peak 4 C-OH | | Peak 5 C-O-C | | Peak 6 C=O | | Peak 7 O-C=O | | O/C ratio % |
|---|---|---|---|---|---|---|---|---|---|---|---|---|---|---|---|
| | BE | % | BE | % | BE | % | BE | % | BE | % | BE | % | BE | % | |
| S1 | 284.4 | 70.8 | 284.8 | 27.9 | - | - | 285.7 | 0.7 | 286.3 | 0.2 | 287.1 | 0.2 | 289.2 | 0.2 | 1.3 |
| S2 | 284.5 | 75.0 | 284.9 | 22.9 | - | - | 285.8 | 1.1 | 286.5 | 0.6 | 287.2 | 0.2 | 289.2 | 0.2 | 2.0 |
| S3 | 284.6 | 83.5 | 285.0 | 13.6 | - | - | 285.9 | 1.8 | 286.3 | 0.6 | 287.2 | 0.4 | 289.2 | 0.2 | 2.9 |
| S4 | 285.0 | 69.4 | 285.4 | 9.8 | 283.7 | 15.3 | 286.3 | 2.4 | 287.0 | 0.8 | 287.7 | 1.8 | 289.6 | 0.4 | 5.5 |

The peak 2 (P2) in the spectra is attributed to $CH_x$ component which is positioned at about 0.4 eV higher BE than that of C-C $sp^3$ component [14]. This peak arises due to the C-H bonds that are present on the diamond surface, grain boundaries and also sub-surface of diamond films. As can be seen from the Table 1, the gradual decrease in the $CH_x$ component is observed upon ozonation indicating the removal of surface hydrogen. Nearly 18 % of $CH_x$ component is decreased from fully hydrogenated to fully ozonated diamond surfaces. However, a significant $CH_x$ content of ~ 10 % is still present on OD even after 1200 s ozanation. This suggests that the $CH_x$ bonds still can present in the sub-surface films.

The peak 4,5,6 and and 7 in C1$s$ spectrum represent the BE of hydroxyl C-OH, epoxy C-O-C, carbonyl C=O and carboxyl O-C=O, respectively [31]. Here, we have estimated the net oxygen concentration solely based on C1s signal which is the sum of the various types of C-C and C-O bondings [32]. The total area of all the C1s peaks is directly proportional to the number of carbon atoms present and the individual area of the each C1s peak is proportional to the number of carbon atoms with the specific functional group. Thus, it is possible to estimate the O/C ratio using the contribution of different species of hydroxyl C-OH (1:1), carbonyl C=O (1:1), epoxy C-O-C (1:2) and carboxyl O-C=O (2:1) with the following equation [32],

$O/C_{ratio} = (A_{C-OH} + A_{C=O} + ½.A_{C-O-C} + 2.A_{O-C=O})/A_{tot}$ …… (1)

where, $A_x$ is the area under the curve of specific chemical bonding and $A_{tot}$ is the total area under the curve of C1s.

As shown in Table 1, the HD sample inherently possesses some amount of oxygen content ~ 1.3 % which is chemisorbed on the surface during exposure to air and it is essential for surface transfer doping [33]. Further, the ozonated samples S2, S3 and S4 possess about 2.0, 2.9 and 5.5 % of oxygen content, respectively. Hence, it is clear that the oxygen concentration on diamond surface increases with ozonation duration. The decrease of $CH_x$ component and increase of oxidized carbon component confirm the simultaneous removal of surface hydrogen and reconstruction of diamond surface with oxygen during ozonation. Note that the estimated O concentration is found to be smaller for partial and fully ozonated diamond surface since the XPS spectra were recorded at the electron polar angle of θ=0 to the surface normal which probes more bulk properties than that of surface [15].

The peak 3 in XPS spectra is associated with the BE of C=C *sp²* bond which is positioned at -1.3 eV with respect to the C-C sp³ component [31]. The C=C *sp²* component is below the detectable limit for the S1, S2 and S3 samples. However, the peak 3 in sample S4 is prominent and very broad with contribution of ~ 15 % of area under the curve. This indicates that the sample S4 consists of significant amount of non-diamond carbon bonds and relatively higher content of oxidized carbon bonds due to the diamond surface oxidation process [15].

## 4. Discussion

Based on the above mentioned observations, it is clear that the chemical structure and physical properties of the HD surface vary significantly with the amount of O- substitution at H-site. In fact, the H- functionalized diamond surface transforms from hydrophobic to hydrophilic nature due to the reduction in van der Waals interactions between the C-H / C-O bonds and water molecules, upon ozonation. On the other hand, the sheet resistance of the HD films increases exponentially when the surface H atoms are replaced by O atoms on diamond surface. The increase in sheet resistance with O-functionalization can be explained as follows: On fully H-functionalized diamond surface, the C-H bonds are polarized with negative charge on C atoms and positive charge on H atoms, as hydrogen exhibits lower electronegativity as compared to carbon. This dipole layer creates a potential step on the surface which pulls the vacuum energy level below the conduction band minimum that results in surface with NEA of ~ -1.3 eV [9]. In HD with NEA, when the atmospheric impurity molecules are adsorbed on diamond surface (chemical potential of water is ~ 4.8 eV which lies below the valence band maximum of diamond), electron transfer takes place spontaneously from diamond valence band to the lowest unoccupied molecular orbital of the adsorbate molecules through electrochemical redox reactions resulting in the hole accumulation on the surface [34]. These holes shift the surface Fermi level position ($E_F$) below the valence band maximum ($E_v$) and give rise to a corresponding upward band bending depending on the charge density. This is well known as surface charge transfer doping process. It may be worth mentioning that while the transfer-doping mechanism is generally accepted, there is some uncertainty regarding the exact electro-chemical process involved in the transfer doping and also, the molecular species which acts as the acceptor on diamond surface [11]. Moreover, when the surface H- atoms are partially substituted by O atoms through ozonation, the local electronic structure changes due to the formation of C-O related bonds. Specifically, it is typically understood that a mixed hydrogen/oxygen surface will show an EA which is a simple average of the constituent EA values. As such, transfer doping is unlikely to occur with significant oxygen coverages on diamond surface [33,35]. Thus, the local EA is tuned by varying H- and O- atoms to control the net hole density on diamond surface, as evidenced by Hall measurements. Consequently, the sheet resistance of the HD films increases drastically with ozonation duration as shown in Fig. 3.

Apart from systematic increase in sheet resistance, the intensity of FL emission of $SiV^{-1}$ color centers increases gradually with ozonation on HD surface as shown in Fig. 5. The surface charge states are known to influence the FL emission of color centers in diamond. For example, the Si-V color centers are present throughout the bulk diamond film with a mixed charge states such as $SiV^+$, $SiV^0$ and $SiV^-$ [36]. The charge state of SiV color center in diamond depends on the Fermi level position with respect to the charge transition level. For SiV centres, the $SiV^{0/-}$ charge transition level is predicted to lie within the forbidden band gap at ~ 0.4 eV above the valence band of diamond [36]. As discussed in electrical measurements, surface Fermi level position ($E_F$) shifts below the VBM ($E_V$) due to surface transfer doping, and a corresponding upward band bending occur in HD surface. This upward band bending from surface into the bulk extends approximately 1-5 nm into the bulk with a formation of 2DHG layer [37]. Thus, $E_F$ position in this few nanometer region can be tuned by manipulating the surface doping density, subsequently to change the charge state of the $SiV^-$ color centers. Based on the existing literature [10], we estimate the $E_F-E_V$ to be of ~ 0.4, 0.5, 0.6, and 1.0 eV for S1, S2, S3, and S4 respectively, which occur due to the lowering of upward band bending in samples S2 and S3 and a downward band bending in sample S4. Hence, the increase of $SiV^-$ emission upon ozonation is due to lowering of upward band bending with $SiV^{0/-}$ charge state transition. However, the increase in intensity of $SiV^-$ emission is not very high, as expected due to large variation in $E_F-E_V$ as a function of ozonation duration. Note that the $SiV^-$ color centres are present in the bulk volume of the diamond films. However, due to surface functionalization / doping, the Fermi level change only occurs at a few nanometer of thickness from surface. Hence, only the $SiV^-$ color centres present on the surface and sub-surface of diamond undergo charge state conversion during surface doping but not in the bulk of the diamond film. The ratio of SiV$^-$ color centres present in the surface and sub-surface to bulk volume of the film is marginal. Hence, we could only observe a slight changes in the fluorescence emission intensity of SiV$^-$ color centers due to ozonation.

The distinct feature in the XPS analysis of the diamonds is that the evolution of peak 3 at the BE of ~ 284 eV and blue shift of peak 1 to higher BE of ~ 285 eV for 1200 s ozonated diamond films. Despite the blue shift of C-C $sp^3$ bonds is well recognized in OD, the origin of peak 3 at the BE of ~ 284 eV is not clearly understood. Generally this peak is assigned to the non-diamond carbon bonds associated $sp^2$ hybridized carbon since the diamond surface undergoes $sp^2$ carbon surface reconstruction during with oxidation process [15]. On the other hand, Seshan et al [38] had reported that the peak with BE of ~ 284 eV (peak 3 in this study) in UV/ozonated diamond film may evolve from C-H bonds that are not completely ozonated in the sub-surface areas and also, it may originate related to surface roughness. Thus, the peak 3 in S4 is either due to C-C $sp^2$ bonding or associated with the unreacted C-H bonds that are present in sub-surface area. We note here that our conventional laboratory XPS analysis without angle-dependent measurements seem to be incomplete study in establishing the direct evidence for C-H bonding. However, the pristine hydrogenated diamond (HD) and partially ozonated HD for 30 and 60 s clearly display a redshift in binding energy (BE) of C-C $sp^3$ bonding while the peak corresponding to C-C $sp^3$ bonding in highly ozonated diamond surface (for 1200 s) does not exhibit any shift in BE. Also, we could observe a clear increase in sheet resistance with corresponding decrease in carrier concentration (macroscopic electrical measurement) and decrease in water wetting contact angle which confirm the successful surface ozonoation process. Hence, we interpret the redshift in BE as the signature of the C-H bonding on diamond surface in support of the existing literature [10,14].

Another important observation by c-AFM study is that an inhomogeneous current conduction on GI of different facets of diamond. Such inhomogeneous current mapping can be associated with the different electronic properties of the diamond planes [10,39,40]. For example, the work function of H-terminated diamond (111) plane is 3.5 eV and it increases upto ~ 5 eV when the surface H-terminated is replaced with O-termination [39]. Similarly, the work function of H-terminated diamond (100) is ~ 4.2 eV and it increases with the coverage of electronegative atoms [40]. In the c-AFM measurements, the AFM tip and diamond surface can be considered as if they are separated by a few angstroms distance with nitrogen molecules or a thin water layer which can act as tunneling medium. Hence, we attribute the higher current flow (~ 4 nA) on (111) planes as compared to very lower current (~ 100 pA) on (100) planes of pristine and 60 s ozonated HDs to the lower work function of HD (111) as compared to HD (100) since a low work function is well known to enhance elecion field emission characteristics of a material. On the other hand, the difference in work function of (111) and (100) planes can be negligible after significant amount of ozonation and hence, the current flow is comparable on both the (111) and (100) planes, as can be seen from Figs. 4(m,n). Although the field emission measurements are performed under large separation distance between the electrodes, this c-AFM studies can also be considered as analogue to tunneling current measurement with a few angstrom separation between the electrodes. However, we accept the opinions which may differ among experimentalists.

## 5. Conclusions

In summary, the growth of polycrystalline hydrogenated diamond (HDs) with good structural quality and high surface conductivity are demonstrated. The surface electronic property is tuned by partial ozonation of HDs. The O- functionalization is qualitatively evaluated by the surface transformation from hydrophobic to hydrophilic nature. Further, the sheet resistance of HDs is found to increase from ~ 8 kΩ/sq. to over 10 GΩ/sq. as a consequence to the reduction of surface carrier density with O- functionalization. Furthermore, the fluorescence emission intensity of $SiV^-$ color centers increases with O termination since the charge state of the color center is tuned by reduction hole density on diamond surface. The surface band bending associated with C-H bonding on pristine and partially ozonated diamond surface is manifested as redshift in BE of C-C $sp^3$ bonding. The uniqueness of the study reports on the local variation of current conduction on the (111) and (100) facets of the polycrystalline HD surface as a function of ozonation duration. Further, the local current decreases drastically on (111) facets with ozonation indicating its higher oxidation rate than that of (100) facets. Also, c-AFM studies confirm the presence of charge transfer doping mechanism on pristine and partially ozonated diamond, and its absence on fully ozonated diamond surface. In conclusion, the surface properties of the hydrogenated polycrystalline diamond can be tuned systematically using partial O-termination. The partial O-termination offer advantages to control the electronic device characteristics of hydrogenated polycrystalline diamond.

**Acknowledgement**

One of the authors, N.M.S, acknowledges IGCAR, DAE, Government of India for research fellowship. Authors thank Arindam Das for providing ozonation facility and fruitful discussion and also, thank anonymous reviewers for their careful reading of our manuscript and their many insightful comments and suggestions.